\documentclass[eqno,twoside,12pt]{article}

\usepackage[dvips]{graphicx}
\usepackage{epsfig}
\usepackage{amsfonts}
\usepackage{amssymb}

\newcommand{\ms}{\medskip}

\newcommand{\ps}{\rightarrow}

\newcommand{\tg}{{\rm tg}}

\def\bbbz{{\mathchoice {\hbox{$\sf\textstyle Z\kern-0.4em Z$}}
{\hbox{$\sf\textstyle Z\kern-0.4em Z$}} 
{\hbox{$\sf\scriptstyle Z\kern-0.3em Z$}} {\hbox{$\sf\scriptscriptstyle
Z\kern-0.2em Z$}}}}
\def\bbbI{{\rm 1\kern-0.25em I}}

\newlength{\au}
\begin{document}
\title {Statistical description of
domains in the Potts model}
\author{
\begin{minipage}{\au} K.~Lukierska-Walasek\\
{\normalsize Institute of Physics\\[-1mm]
University of Zielona G\'ora\\[-1mm] 
ul. Z. Szafrana 4a\\[-1mm]
65-516~Zielona~G\'ora,~Poland}
\end{minipage}\and 
\begin{minipage}{\au} K.~Topolski\\
{\normalsize Institute of Mathematics\\[-1mm] 
Wroc\l aw University\\[-1mm]
Pl.~Grunwaldzki 2/4\\[-1mm]
50-384 Wroc\l aw, Poland
}
\end{minipage}}
\date{}
\maketitle

\begin{abstract}
The Zipf power law and its connection with the inhomogeneity of the
system is investigated. We describe the statistical distributions of the
domain masses in the Potts model near the temperature-induced phase transition.
We found that the statistical distribution near the critical point is described by
 the power law form with  a long tail, while beyond the critical 
point the power law tail is suppressed.
\end{abstract}

\vspace{0.8cm}

\noindent
 We use the Potts model \cite{Potts} for the description of the phase transition.
The Potts model is the generalization of the Ising model  with more than two spin components
and it has more experimental realizations than  the Ising model. For a  detailed review of Potts model see \cite{Wu}.

The  Hamiltonian for the $q-$state Potts model \cite{Wu} is:

\vspace{-3mm}
\begin{equation}
 H=-\sum\limits_{ij} J_{ij}\delta_{\sigma_i,\sigma_j},
\end{equation}
where $\sigma_{i}\in\{1,2,...,q\},$ $\delta_{x,y}$ is the Kronecker delta
$$\delta _{x,y}=\left\{
\begin{array}{l}
1 \quad \mbox{if \ $x=y$,}\\[2mm]
0 \quad \mbox{otherwise}.
\end{array}\right .
$$
and 
$$
J_{ij}=\left\{
\begin{array}{l}
J \quad \mbox{if \ $i,j$ are neighbour pairs of spins,}\\[2mm]
0 \quad \mbox{in opposite case}.
\end{array}\right .
$$
The case $q=2$ describes  the Ising model.

The results presented hereafter are  obtained by applying the Monte Carlo techniques, 
based on the Metropolis algorithm \cite{Metropolis}, to the two-dimensional 
$q-$state Potts model with the periodic boundary conditions and with $q=3$ and $6$.

A number of exact results for the two -dimensional Potts models are known in the infinite volume limit. For example, the phase transition appears at the critical temperature $\,T=T_c\,$ 
($\,T_c=2J/k_B\,\ln(1+\sqrt{q})\,$).
It is the second order phase transition for $q\leq 4$ and the first order one for $q\geq 5$, see \cite{Baxter} and \cite{Wu}.

The main goal of this paper is the statistical description of the domains in 
Potts model  when one approaches the critical point of the phase transition
induced by the temperature.  This problem has been investigated for the 
Ising model in \cite{LT}.
Our considerations concerning the statistical description of domain masses have 
universal character and may be used to arbitrary fractal system of elements 
which are described by the random variables $x$. This variables can be 
listed in the decreasing order. For example: the rank of the city 
connected with its population, the frequency of the occurrence of 
any word in the text, the trading value of largest European's 
companies or the hyperbolic processes in finance \cite{Bouch}, \cite{Bibby}.\\ 
We consider the Zipf power law \cite{Zipf} in Bouchaud notation \cite{Bouch}:
\begin{equation}
x=k^{-\frac{1}{\mu}}
\end{equation}
This law is used often in the description of  self-organized critical phenomena.

In our case $\,x\,$ is the number of the Potts spins in the domain called the domain 
mass and $\,k\,$ denotes the rank of the domain mass $\,x$.The geometrical clusters are considered \cite{Stauffer}. 
The greatest cluster has the rank 1, smaller 2 and so on.
In the Potts model, clusters are sets of nearest neighbour sites occupied by Potts spins
 with the same orientation. 
For a given configuration of spins clusters are determined uniquely.
There is the well establish connection between thermal and geometric phase transitions.
 In the dimension two an infinite cluster appears exactly at the critical point.
The $q-$state Potts model
has the ground state degeneracy $q$ and after a quench from the high-temperature phase,
small domains start to grow, thus reducing the domain boundary curvature. 
One can notice that there is a difference between the domains of the Potts model with $q=2$ 
(the Ising model case) and the Potts model with $q>2$.
For $q=2$ the domain boundaries are represented by long {\em filaments} of one phase within 
the other one, whereas for $q>2$, the domains resemble the structure characteristic for 
polycrystalline grains \cite{boundary1}, \cite{boundary2} and \cite{boundary3} 
 - i.e. the boundaries are straight and meet at fixed angles. 
 
We take into account, in our considerations,  all domains and we concentrate on their size.
 Bouchaud \cite{Bouch} pointed out the strong correlation between the Zipf power
law and the inhomogeneity of the system.
We are going to test the log-log distribution of domain masses versus the rank index $k$ when we come near the critical point in the temperature-induced  phase transition.

The slope $\alpha$ of the regression line
describing relation between the logarithm of the cluster mass and the logarithm of its rank
 is given by$\mu$ $(-\frac{1}{\mu} =\tg\,\alpha)$, and it characterizes the inhomogeneity
of the physical structure of the system. The inhomogeneity of the system means that
its structure has become  fractal and more hierarchical. We expect that $\mu$ value will depend on whether we are far or close to the critical point. In particular one should have $\mu\approx 1$ in the critical point and $\mu>1$ beyond the criticality.
This follows from the following considerations.
It is shown in \cite{Stanley} that there is a conjecture between Zipf exponent and the Hurst exponent $H$ \cite{Hurst} of the form
\begin{equation}
1/\mu = |2H - 1|
\end{equation}
It is well known that any complex system exhibits long-range (infinite-long) correlations at the critical point. This corresponds to the biggest value of Hurst exponent, i.e. $H = 1$. In such a case one obtains from Eq.(3) $\mu\approx 1$.\\
The criticality in the rank statistics was also considered experimentally in \cite{Moz1},\cite{Moz2} investigating distributions of island areas of discontinuous metal films near the percolation threshold.\\
In our consideration we shall concentrate  on  random
variable $X$ which we shall call $\,[\mu]-variable.$ We say that the random variable $X$
is $[\mu]-variable$ if for some $x_0>0$ and $\mu>0,$ a tail of its distribution
function decays as $\left(\frac{x_0}{x}\right)^\mu.$

The main property of $\,[\mu]-variable$ is that all its moments
$\,m_q=<x^q>\,$ with $\,q\geq\mu\,$ are infinite.
The distribution of the cluster size seems to be 
$[\mu]$-distribution and the index $\mu$ 
is a critical exponent.

The simplest technique which helps to test heavy tail hypothesis and estimates
the tail index $\mu$ is based on the following reasoning, (for more details see \cite{Kratz}).

Let $X_1,\,X_2,\,...,\,X_n$ be independent and identically distributed 
random variables with  distribution function $F$.
We may re-arrange the $X_i$ in the decreasing order. Denoting the largest  value, with rank 1,
 by $X_{(1)},$ the next largest, with rank 2, by $X_{(2)}$ etc., we arrive 
at the sequence of order statistics
 $$X_{(1)}\geq X_{(2)}\geq\,...\geq X_{(n)}.$$

Assume that $X$ is a $\,[\mu]-variable$ with tail of its distribution function  satisfying
for some $x_0>0$ and $\mu>0$ 
\begin{equation}\label{pareto}
 P(X>x)=\left(\frac{x_0}{x}\right)^{\mu},\quad x>x_0.
\end{equation}

The assumption (\ref{pareto}) means that the distribution function of $\frac{X}{x_0}$ 
is the Pareto distribution with the left endpoint 1 and the parameter $\mu$, and for $x>0$
$$P\left(\mu \ln \frac{X}{x_0}>x\right)=\exp(-x)$$
which means that the  distribution function of the random variable $\mu\ln
\frac{X}{x_0}$ is exponential with parameter 1.
\newpage
Therefore
$$P(\ln\,X>x)=\exp\left(-\frac{x-\ln x_0}{\mu^{-1}}\right),$$
which is a tail of shifted by $\ln x_0$ and rescaled by $\mu^{-1}$ exponential  
distribution function with parameter 1.

Drawing  the quantile to quantile plot for the exponential distribution function 
we conclude that
$$\left\{\left(-\ln \left(\frac{i}{n+1}\right ),\,\ln X_{(i)}\right),\,1\leq i\leq n\right \},$$
should be linear with slope $\mu^{-1}$ and intercept $\ln x_0$.
In this way we obtain the Zipf power law.
We may estimate $\mu$ in a different way.
The most popular estimator of $\mu$ is the   Hill estimator, \cite{Hill}.
For the convenience of the reader we recall its definition.
The Hill's estimator of $\frac{1}{\mu}$ based on $m$ upper-ordered statistics 
$X_{(1)}\geq X_{(2)}\geq...\geq X_{(m)}$ is defined as
\begin{equation}\label{hill}
H_{m,n}=\frac{1}{m}\sum_{i=1}^{m}\log\frac{X_{(i)}}{X_{(m+1)}}.
\end{equation}

Our simulations were performed on  two dimensional lattices $600\times 600$ with periodic
boundary conditions.
We used the Metropolis algorithm \cite{Metropolis}. All simulations were initiated with the random initial 
orientation of spins. We thermalized the system to equilibrate the orientations of spins. The number of Monte Carlo 
steps which are necessary to reach the equilibrium state was chosen by measuring the average energy of one 
spin.

We performed 1 000 000 Monte Carlo steps to equilibrate the system. 
Once  the equilibrium of the system is reached  we took sample of the cluster configuration.
Then we studied the distribution of the cluster masses. The total mass of the cluster was defined as the number of spins in the cluster. 

Our aim is to connect the statistics of domain masses with the way the critical point is reached.
When we start to advance from the
paramagnetic phase to the critical point $\,(T\ps T_c)\,$ (as it is  seen in Fig. 1, 
for $q=3$, and Fig. 2 for $q=6$) 
the slope of  straight lines representing (in log-log scale) the
Zipf power law  increases when $\mu>1$. 
\newpage

\begin{figure}[ht]
\begin{center}
\includegraphics[height=50mm, width=10cm]{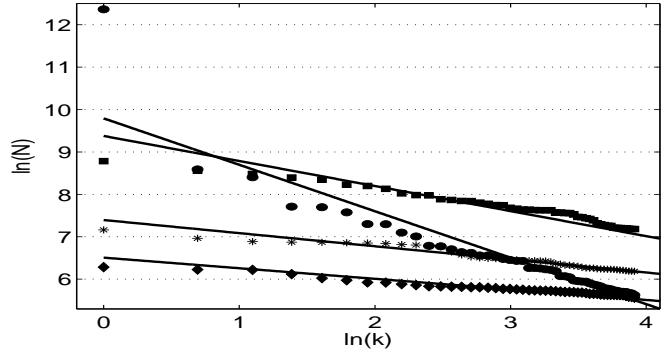}
\caption{{\small
The log-log distribution of the domain masses $x_k$ versus the rank order
index $k$ for the 3-state Potts model for $\beta=0.40$ ($\Diamond$), $0.45$ ($\ast$), $0.49$ 
($\Box$) and critical $\beta$ ($\circ$), for one configuration. The straight line are with the slopes $-1/\mu$ given in Table 1.}}
\end{center}
\end{figure}

\bigskip

\begin{figure}[ht]
\begin{center}
\includegraphics[height=50mm, width=10cm]{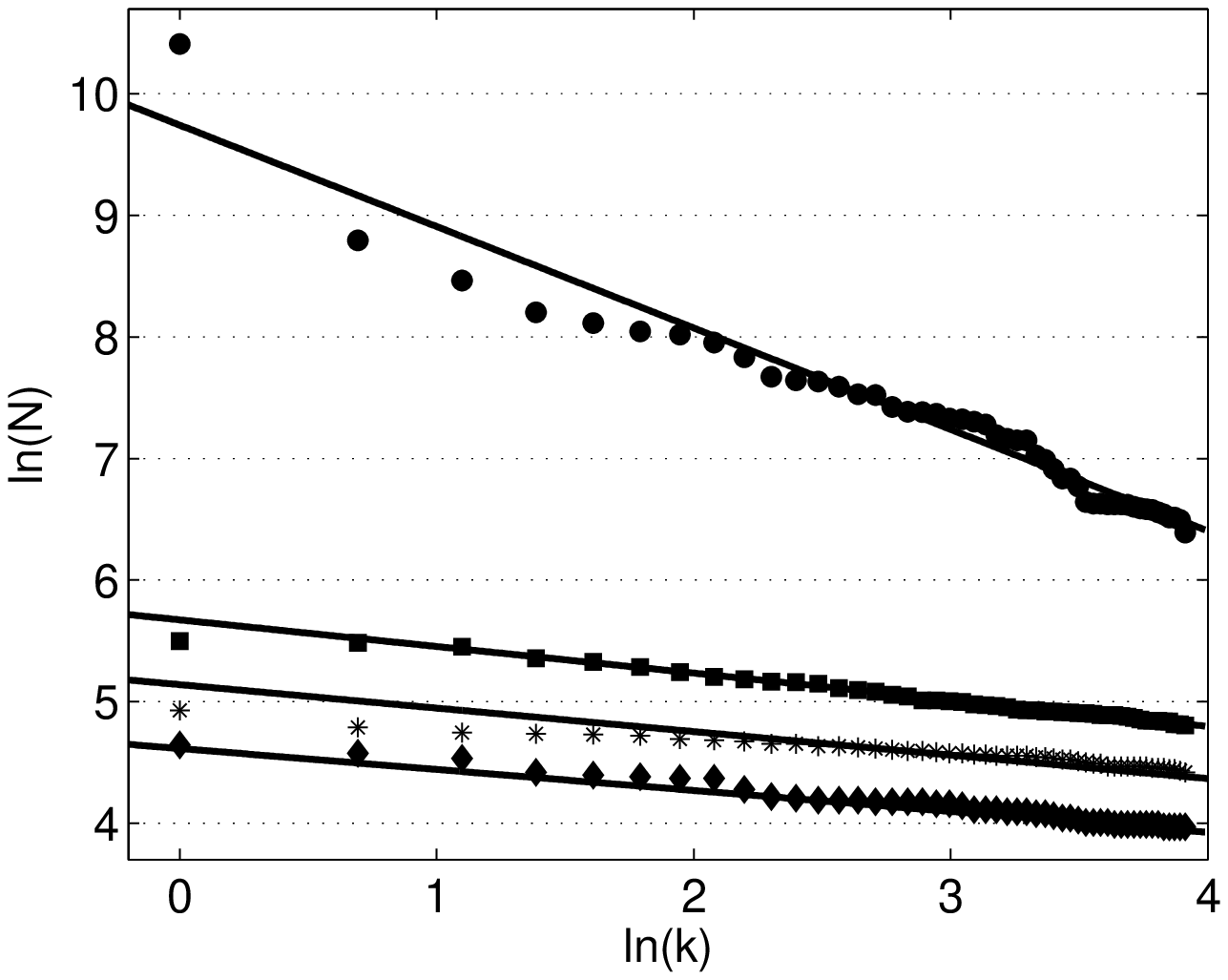}
\caption{{\small
The log-log distribution of the domain masses $x_k$ versus the rank order
index $k$ for the 6-state Potts model for $\beta=0.50$ ($\Diamond$), $0.55$ ($\ast$), $0.59$ 
($\Box$) and critical $\beta$ ($\circ$), for one configuration. The straight line are with the slopes $-1/\mu$ given in Table 2.}}
\end{center}
\end{figure}

\newpage
The estimations of indexes $\frac{1}{\mu}$, for the 3-state Potts model with different inverse temperature $\beta = 1/T$,  based on 1000 realizations for $600 \times 600$  lattice are presented in Table 1 and Fig.3. 

\medskip

\begin{center}
\centerline{Table 1. {\small The estimation of indexes $\frac{1}{\mu}$ with standard deviations $s$.}} 

\medskip

\begin{tabular}{|l|l|l|}
\hline
$\beta$ & $\mu^{-1}$ & $s$ \\ \hline
$0.40$ & $  0.24949$ & $0.023768 $ \\
$0.45$ & $ 0.30957$ &$0.030429$\\ 
$0.49$ & $0.59078$ & $0.046805 $\\  
$\beta_{kr}\simeq 0.5025,\quad $ & $1.09547$ & $0.047967 $\\ 
\hline
\end{tabular}
\end{center}

\bigskip

\begin{figure}[ht]
\begin{center}
\includegraphics[height=6cm, width=8cm]{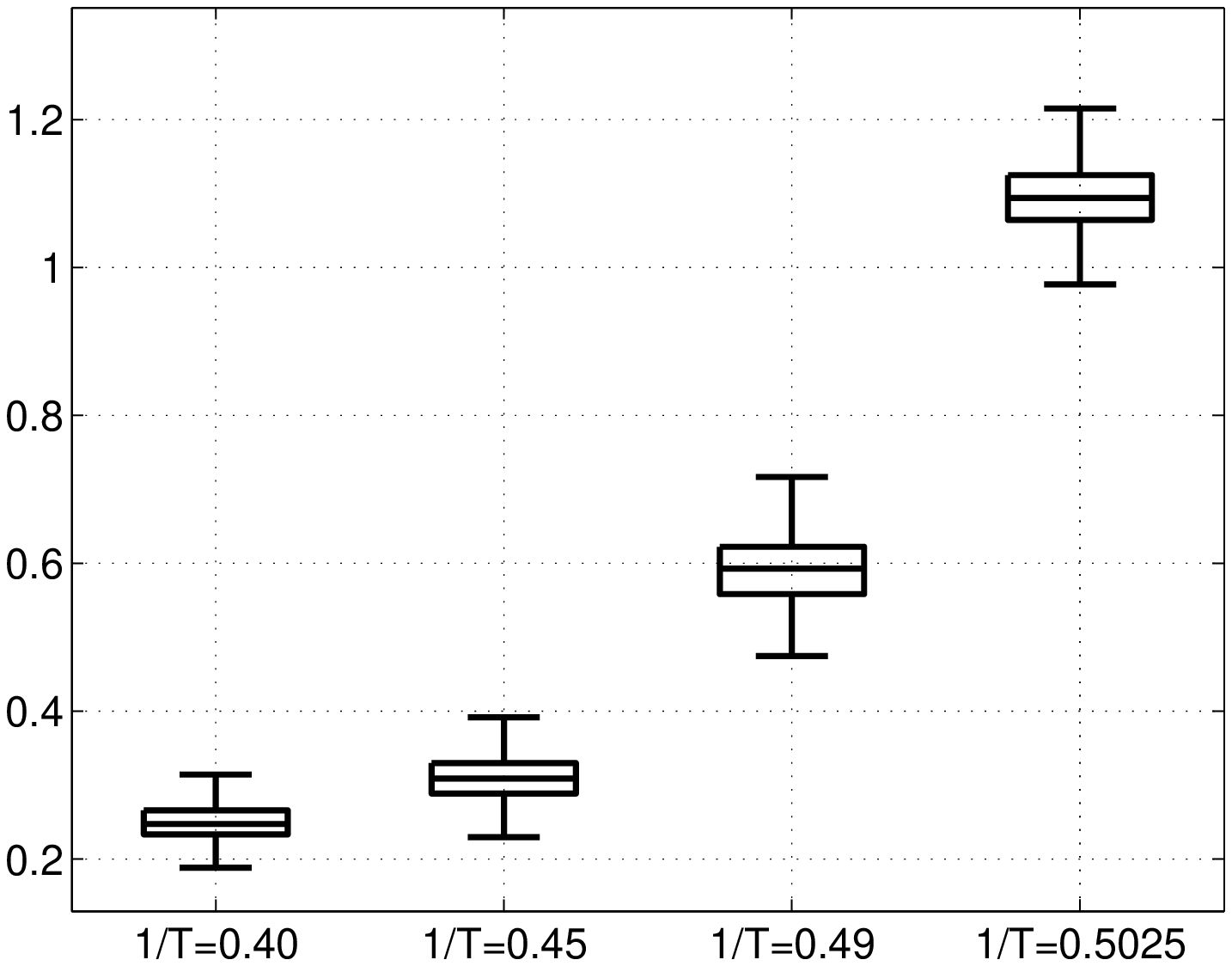}
\caption{{\small Box plot: The mean value of $\mu^{-1}$ (--) with standard deviation and
95\% confidence interval, for inverse temperatures $\beta=0.40, 0.45, 0.49$ and and close to critical $0.50253$, obtained by averaging of 1000 Monte Carlo realizations of the 3-state Potts model on  $600\times 600$ lattice.}}
\end{center}
\end{figure}
\newpage
\noindent
The same estimations of indexes $\frac{1}{\mu}$ and its standard deviations as in Table 1 and Fig 3, but for 6-state Potts model,
are presented in Table 2 and Fig.4.
\medskip

\centerline{Table 2. {\small The estimation of index $\frac{1}{\mu}$  with standard deviation $s$.} }

\medskip

\begin{center}
\begin{tabular}{|l|l|l|}
\hline
$\beta$ & $\mu^{-1}$ & $s$ \\ \hline
$0.50$  & $ 0.172505$ & $0.017357$\\
$0.55$  & $0.193189$ & $0.020260$\\ 
$0.59$ & $ 0.219017$ & $0.022300$\\ 
$\beta_{kr}\simeq 0.6191 \quad$ &$ 0.834464$ &$0.043940$ \\
\hline
\end{tabular}
\end{center}

\bigskip

\begin{figure}[ht]
\begin{center}
\includegraphics[height=6cm, width=8cm]{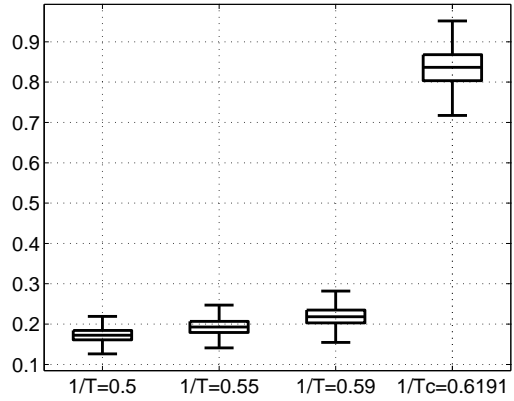}
\caption{{\small
Box plot: The mean value of $\mu^{-1}$ (--) with standard deviation  and
95\% confidence interval, for inverse temperatures $\beta=0.50, 0.55, 0.59$ and close to critical $0.6191$, obtained by averaging of 1000 Monte Carlo realizations of the 6-state Potts model on  $600\times 600$ lattice.}}
\end{center}
\end{figure}

\newpage
\noindent
Fig. 5 and Fig. 6 represent the dependence between the number of domains with mass x normalized by number of configurations
 and mass(x) for 3-state and 6-state Potts models respectively, near the critical point.
The distribution of domain mass is $F[\mu]$-distribution  with
$\mu\approx 1$ ,which means that it satisfies power law with a long tail: $\frac{x_0^{\mu}}{x^{1+\mu}}$,where $x_0$ denotes a typical scale and $\mu\simeq 1$.

\begin{figure}[ht]
\begin{center}
\includegraphics[height=6cm, width=8cm]{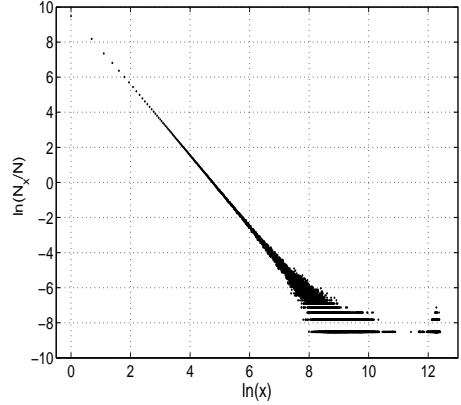}
\caption{{\small The dependence between the  number of domains with mass x normalized by number of configurations
 $(N_x/N)$ and the mass $(x)$
in the log-log scale for the 3-state Potts model near the critical
point $T_c$ for 5000 configurations and  $L=600$.}}
\end{center}
\end{figure}

\begin{figure}[ht]
\begin{center}
\includegraphics[height=6cm, width=8cm]{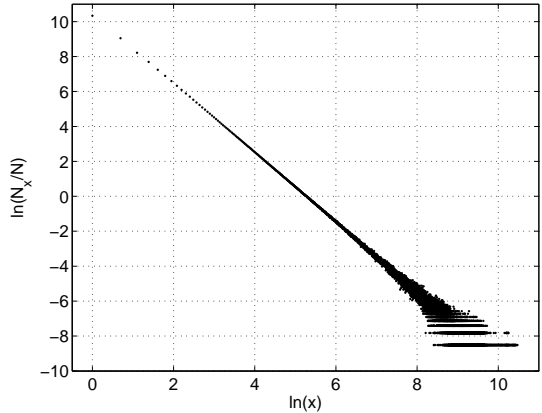}
\caption{{\small The dependence between the  number of domains with mass x normalized by number of configurations
 $(N_x/N)$ and the mass $(x)$
in the log-log scale for the 6-state Potts model near the critical
point $T_c$ for 5000 configuration and  $L=600$.}}
\end{center}
\end{figure}
\noindent
Fig. 7 and Fig. 8 represent the dependence between the  number of domains with mass x normalized by number of configurations
 $(N_x/N)$ and the mass $(x)$  of domain masses for 3-state and 6-state
Potts models beyond the critical
region when $\mu>1$,  respectively. In this case the distribution of domain masses
is without heavy power-law tail.
This results are in the agreement with the standard percolation theory \cite{Stauffer}.

\begin{figure}[ht]
\begin{center}
\includegraphics[height=6cm, width=8cm]{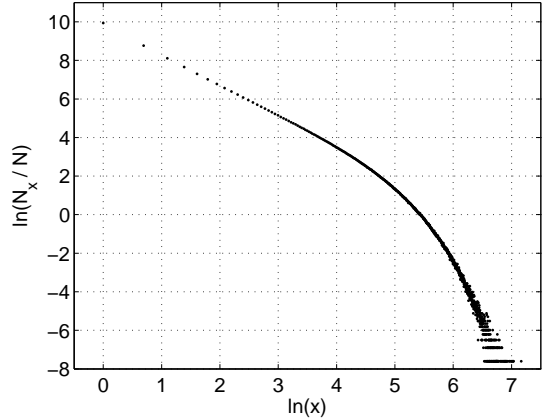}
\caption{{\small The dependence between the  number of domains with mass x normalized by number of configurations
 $(N_x/N)$ and the mass $(x)$
in the log-log scale for the 3-state Potts model for $\beta=0.4$ for 2000 configurations and  $L=600$.}}
\end{center}
\end{figure}
\begin{figure}[ht]
\begin{center}
\includegraphics[height=6cm, width=8cm]{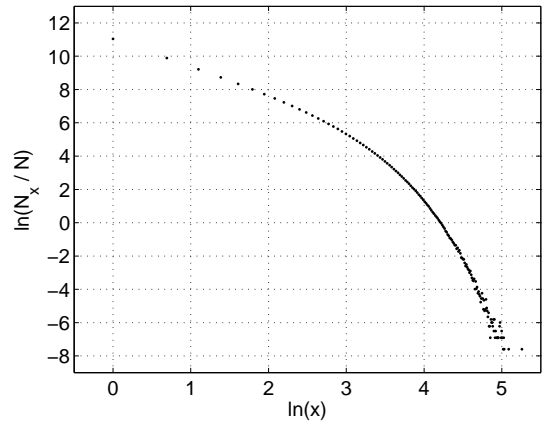}
\caption{{\small The dependence between the  number of domains with mass x normalized by number of configurations
 $(N_x/N)$ and the mass $(x)$
in the log-log scale for the 6-state Potts model for $\beta=0.5$ for 2000 configuration and  $L=600$.}}
\end{center}
\end{figure}

\ms
The numerical results presented in Figures 5-8, usually can be deduced from the
 standard percolation theory \cite{Stauffer} and are in the agreement with the previous
 results of \cite{J1} and \cite{J2}. The origin of the power-law is well understood
 and explained as a consequence of the scale invariance at the criticality regime
 due to the divergence of the correlation length. Beyond the criticality range, where
 the system is a fractal its correlation length is finite so the power-law tail
 in the distribution is suppressed.
One of the possible experimental confirmations of the Zipf power
 law is the distribution of island areas of discontinuous metal films,
 obtained by  evaporation  of  dielectric substances \cite{Moz1}, \cite{Moz2}. Another one 
is the distribution of local fields intensities \cite{Lib}. Unfortunately, the results of analogous experiment on ferromagnetic are not known.

The main result of our  paper is obtained by applying the Zipf power
 law to the criticality region and beyond it for the phase transition in the Potts model.

Further, we have shown, by the simulation results, that the exponent  in the Zipf
 power law describes the long tail 
behaviour of the distribution of  domains masses at the critical point when $\mu\approx 1$.

These results deal with the physical phenomena different from those
 considered e.g., in \cite{Moz1}, \cite{Moz2} and \cite{Lib}. However in all
 cases the formulas describing properties of the system are
 similar and they are based on the Zipf power law.
The statistical description gives the opportunity to test
 the phase transition not only in the critical region but
 also beyond the criticality.\\[3mm]
 
{\bf Acknowledgments.}
One of the author (K.~L-W) would like to express her gratitude to dr Dariusz Grech for the interest in this paper and valuable comments.


\begin{thebibliography}{99}

\bibitem{Potts} R. B. Potts, Ph. D. Thesis, University of Oxford (1951),\\
Proc. Camb. Phil. Soc. 48 (1952) 106.

\bibitem{Wu} F. Y. Wu, Rev. Mod. Phys. 54 (182) 235.

\bibitem{Metropolis} N. Metropolis, A. W. Rosenbluth,  M. N. Rosenbluth, A. H. Teller,
 E. Teller, J. Chem. Phys. 21 (1953) 1087. 

\bibitem{Baxter} R. J. Baxter, J. Phys. C 8 (1973) L445.

\bibitem{LT} K.~Lukierska-Walasek, K.~Topolski, Preprint arXiv:~cond-mat.stat-mech 0804.3522v1. 

\bibitem{Bouch} J.P. Bouchand, Proc. Int. Workshop on L\'evy Flights and Related
Topics in Physics (Nice, France, 27--30 June 1994) ed. by M.F. Shlesinger,
G.M. Zas{\l}awsky, V. Frisch, Berlin, Springer, 1995.

\bibitem{Bibby} M. Bibby, M. Sorensen, Handbook of Heavy Tailed Distributions 
in Finance Vol. 1 ed. S. T. Rachev, Elsevier (2003).

\bibitem{Zipf} Human Behavior and the Principle of Least Effort (Addison-Wesley, New York, 1949).

\bibitem{Stauffer} D. Stauffer, A. Aharony {\em Introduction to percolation theory,}
 Taylor and Francis,
London and Philadelphia, 1994.
\bibitem{boundary1} J. M. Lifshitz, Sov. Phys. JETP 15 (1962) 939.

\bibitem{boundary2} G. S. Grest, M.P. Anderson and J.P. Srolovitz, Phys. Rev. B 38 (1988) 4752.
 
\bibitem{boundary3}  M. R. Dudek, J. F. Grouyet, M. Kolb, Surface Science 401 (1998) 220.

\bibitem{Stanley} A. Czirok, R.Mantegna, S. Havlin, H. E. Stanley, Phys. Rev. E.52 (1995) 446.

\bibitem{Hurst} H. E. Hurst, Trans. Am. Soc. Civ. Eng. 116 (1951) 770

\bibitem{Moz1} E. Dobierzewska-Mozrzymas, P. Bieganski, E. Piecul and J. Wojcik,
 J. Phys. Condens Matter 11 (1999) 5561.

\bibitem{Moz2}E.Dobierzewska-Mozrzymas,G.Szymczak,P.Biegañski,E.Piecul,Physica B337(2003)79, 

\bibitem{Kratz} M. Kratz, S.I. Resnick, Stochastic Models, 12 (1996) 699.\\
J.Beirlant, Y. Goegebeeur, J. Teugels and J. Segers, {\em Statistics of 
Extremes, Theory and Applications,} Wiley, Chichester, UK, 2004.

\bibitem{Hill} B. M. Hill, Ann. Stat. 3 (1975) 1163.

\bibitem{J1} W. Janke and A. M. J. Schakel, Phys. Rev. E 71  (2005) 036703.

\bibitem{J2} W. Janke and A. M. J. Schakel, Braz. J. Phys. 36  (2006) 708.
 
\bibitem{Lib} S. Liberman, F. Browers, P. Gadenne, Physica B 279 (2000) 56.

\end{thebibliography}
\end{document}